\documentclass[12pt]{iopart}

\usepackage{amssymb}
 \usepackage{amsfonts}
\usepackage{graphicx}
 \usepackage{mathbbol,fixmath}
 \newcommand{\bj}{{\mathbold j}}
 \newcommand{\bbr}{{\mathbold r}}
 \newcommand{\bv}{{\mathbold v}}
 \newcommand{\bB}{{\mathbold B}}
 \newcommand{\bE}{{\mathbold E}}
 \newcommand{\bJ}{{\mathbold J}}
 \newcommand{\bR}{{\mathbold R}}
 \newcommand{\bbeta}{\mathbold{\beta}}

\begin{document}

\title[Generalization of the Van Cittert--Zernike
theorem]{Generalization of the Van Cittert--Zernike theorem: observers 
  moving with respect to sources}

\author{Daniel Braun$^{1}$, Younes Monjid$^2$, Bernard Roug\'e$^2$ and Yann Kerr$^2$}
\address{
$^{1}$Institut f\"ur theoretische Physik, Universit\"at T\"{u}bingen,
72076 T\"ubingen, Germany\\
$^2$CESBIO, 18 av. Edouard Belin, 31401 Toulouse, France
}

\begin{abstract}
The use of the Van Cittert--Zernike theorem for the 
formulation of the 
visibility function in 
satellite-based Earth observation with passive radiometers 
does not take into account the relative motion
of the observer (the satellite antenna) with respect to sources of
the electro-magnetic fields at the surface of the Earth. The motion of
the observer leads on the one 
hand to a more complex signal due to a pixel-dependent Doppler shift
that is neglected in the standard derivation of the
Van Cittert--Zernike theorem, 
but on the other hand one may hope that it could be employed for a temporal aperture
synthesis, where virtual baselines are created through the motion of
the satellite.   
Here, we generalize the formulation of the aperture synthesis
concept 
to the case of observers moving with
respect to the sources, and to the correlation of fields measured at
times that differ by the travel time of the observer along a
virtual baseline. Our derivation is based on first principles,
starting with the wave propagation in 
the Earth reference frame of electro-magnetic fields arising from
incoherent current sources, and Lorentz transforming the fields into
the reference frame of the satellite. 
Our detailed study leads to the remarkable conclusion that the delay
time due to observer
motion cancels exactly the Doppler effect. This justifies
the neglect of the Doppler effect in existing imaging
systems based on the standard Van Cittert--Zernike theorem. 
\end{abstract}

%
%
%
%
%
\section{Introduction}
Passive microwave remote sensing has a long track record especially in
Earth observation. Currently, there is an increasing interest
in using microwave remote sensing for monitoring key parameters of our
environment. It offers an all-weather capability that is paramount
to understanding and monitoring surface variables and
parameters of Earth. Among the parameters which can be measured are sea surface
temperature \cite{SST}, ocean salinity \cite{SSS_NASA}, soil moisture  \cite{SM} and sea
ice concentration \cite{Sea_ice,Sea_ice_2} to name but a few. For several measurements,
notably sea surface salinity or surface soil moisture, low
frequencies are to be preferred \cite{SM_mapping,surf_param_inversion} which leads to using the
protected frequency bands (e.g., 1400-1427 MHz). However, spatial
resolution then becomes an issue because an adequate resolution
(10 -- 40 km) requires a large antenna (respectively 32 to 8 meters, roughly). As
the launcher's shroud is limited, embarking an 8 meter antenna is
already a challenge which has been addressed differently according to
the science objectives.\\ 

The SMOS mission was the first attempt ever to have an L Band
radiometer on a satellite (the SkyLab mission was an attempt to check
feasibility during a very short period of time in the 1970s \cite{Skylab}). To
address both sea surface salinity and soil moisture it was proposed to
ESA to use spatial aperture synthesis interferometry \cite{Syn_apert}, giving a relatively high spatial
resolution (ranging from 27 to 60 km) with a deployable antenna. The
radiometric resolution is relatively low 
but can easily be compensated by
multi-angular acquisitions \cite{SMOS}. The satellite was
launched in 2009 and is still operational providing good results on
both oceans and land. Alternative approaches
use  filled antennae and do not achieve the same spatial resolution,
such as the Aquarius mission (launched 2011, $\sim$100 km resolution,
\cite{Aquarius}), and the SMAP mission (launched in January,
2015, 
47$\times 61$km resolution \cite{SMAP}).
\\  

Spatial aperture synthesis is a widespread technique in
radio-astronomy \cite{Interferometry_Synthesis_Radio_Astronomy}. It is based on 
precisely timing the arrival times of signals from a given source
at spatially 
separated antennae and then correlating them. 
In the simplest case of negligible motion of the observer relative to
the source and correlation of simultaneously registered signals the 
correlation functions are related by a simply Fourier-type law to the
intensity distribution of the distant incoherent source through the Van
Cittert--Zernike theorem (VCZT) (see eq.(\ref{vCZ}) below). 

The standard derivations of the theorem do not take the
motion of the observer relative to the sources into account.  In
particular, the Doppler shift is neglected. Secondly, one may want to
generalize the theorem to correlations of electric fields measured at
{\em different} times in order to achieve a {\em temporal
aperture synthesis}, where in addition to the physical baselines
provided by different antennae, virtual baselines are created through the
motion of the same antenna.  In astronomy, rotational
synthesis using the motion of Earth has long been used, starting from
radio-astronomy in the 1950s till its advance into the visible in the last
decade
\cite{w._n._christiansen_radiotelescopes_1969,duev_radioastron_2015}.
Satellite-based temporal aperture 
synthesis for radio astronomy is currently being tested on board of the
mission Spektr-R (Radio-Astron) \cite{RadioAstron}, whose measured signals are combined with
those of antennae on Earth and allow one to create virtual baselines
of up to 350,000 km lengths.  The notion of measurements at different
times arises here due to the different arrival times of a signal from
a given small
source.  Also in microwave remote sensing of 
Earth the use of motion of the satellite was
proposed in the recent past
\cite{camps_two-dimensional_2001,park_rectangular_2008} with the same
idea of delays specific for each antenna and pixel. 
    
The purpose of this paper is two-fold: Firstly, we provide the appropriate
generalizations of the Van
Cittert--Zernike theorem  to the  case of an observer moving   relative
to the sources.  Secondly, we
investigate correlations of fields measured at different times,
delayed by the motion of the observer along the desired virtual base
line. These delay times depend on the motion of the
antennae, but not on the positions of the sources.  The question is
whether with such global, pixel-independent shifts virtual baselines
can be created.
Our derivation is based on first principles, using only Maxwell's
equations and a Lorentz transformation from the Earth frame (where the
sources of the electro-magnetic fields are at rest) to the satellite
frame, taken to be moving uniformly with constant speed relative to Earth. 
We show that there is a remarkable cancellation of the phase due to
the first order Doppler effect and the phase due to the virtual
baseline.  This renders the idea of temporal aperture synthesis
through direct correlation of electric fields shifted in time by the
travel time of the satellite over the distance of the virtual baseline
impossible.    At the same time, our generalized derivation shows that
the VCZT in its standard form is still valid to first order in the
speed of the observer and provides therefore a {\em a posteriori}
justification of the use of the theorem for satellite observations of
Earth without including the Doppler shift.

\section{Doppler effect}
Consider two antennae on-board a satellite flying at a height $h$ with
a speed $v_{s}$ in the along-track ($x$-direction) over a current
sources at Earth surface. Let
$\mathcal{R}=(O,\hat{e}_{1},\hat{e}_{2},\hat{e}_{3})$ be the reference
frame fixed with respect to Earth and
$\mathcal{R}'=(O',\hat{e}'_{1},\hat{e}'_{2},\hat{e}'_{3})$ the
reference frame moving with speed $\bv_s$, in which
the satellite is thus at rest with the antennae at fixed positions
$\bbr'$. At time 
$t=t'=0$, the two origins $O,O'$ are taken to coincide. 
Let $\bj(\bbr,t)$ be a current density expressed at the
position $\bbr=(x_{1},x_{2},x_{3})$ in the reference frame
$\mathcal{R}$ relative to Earth. 
One is interested in calculating electric fields at points
$\bbr'=(x_{1}',x_{2}',x_{3}')$ relative to $\mathcal{R}'$,
and with time stamps $t'$,  the time
measured by a clock on board of the satellite. 
\begin{figure}[H]
\centering
\includegraphics[width=10cm]{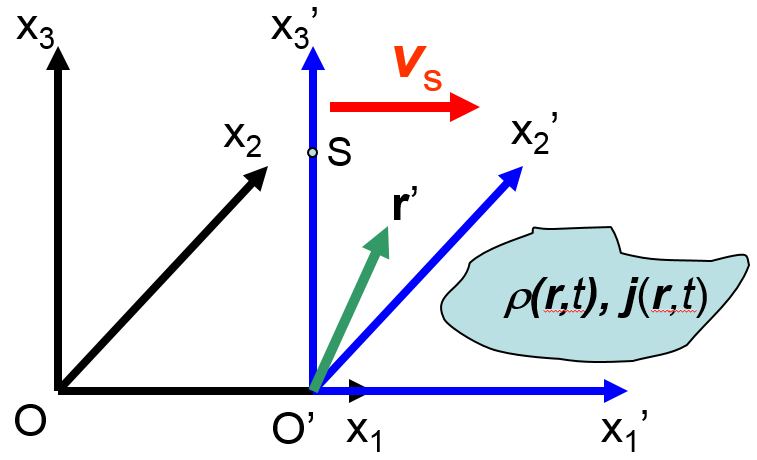}\\
\caption{Earth frame ${\cal R}=(O,\hat{e}_1,\hat{e}_2,\hat{e}_3)$ and
 reference frame ${\cal 
 R'}=(O',\hat{e}'_1,\hat{e}'_2,\hat{e}'_3)$
moving with speed $\bv_{s}$ relative to ${\cal R}$.  Fluctuating
charge densities $\rho(\bbr,t)$ and current densities $\bj(\bbr,t)$
create electric 
and magnetic fields.  Antennae at fixed
positions $\bbr'_i=(x_i',y_i',h)$ with respect to ${\cal R'}$ measure
the electric fields at the proper time $t'$ of the satellite $S$. 
}
\label{fig.frames}
\end{figure}
Two types of
baselines are created when considering two antennae at fixed positions
$\bbr'_1$, $\bbr'_2$ with respect to 
${\cal R}'$: A real
physical baseline between the two antennae and a virtual baseline due
to the displacement of the antenna. The first one, exploited in
spatial aperture synthesis and implemented in SMOS, uses fields
measured at space-time points 
$(\bbr'_1,t')$ and $(\bbr'_2,t')$, implying a physical baseline
$\bbr_1'-\bbr_2'$.    The second one uses measurements recorded in
space-time points
$(\bbr'_1,t_1')$ and  $(\bbr'_2,t_2')$ for two different times $t_1'$,
$t_2'$, which implies a virtual baseline in frame ${\cal R}$ with an
additional ``virtual'' component $\simeq v_s(t_2'-t_1')$ in the
direction of the motion of the 
satellite (neglecting here for clarity the relativistic correction,
see below for the 
precise transformation).  This is therefore a ``temporal'' aperture
synthesis. \\  
We now calculate the electric fields measured by the satellite that
are created by charge and current densities in the
Earth-fixed frame ${\cal R}$.


\subsection{Lorentz transformation}
The sources of the e.m.~fields are microscopic charges and
currents at
the surface of Earth's surface, in thermal equilibrium at some
position dependent temperature $T$. They give rise to an electric field
$\bE(\bbr,t)$, and a magnetic induction
$\bB(\bbr,t)$, originally expressed in the
frame $\mathcal{R}$ fixed to Earth. These fields are 
observed at a space-time point $(\bbr',t')$ in the frame
$\mathcal{R}'$ relative 
to the satellite.  The fields $\bE'(\bbr',t')$ measured
in the satellite frame ${\cal 
  R}'$ can be obtained from a Lorentz
transformation (LT) that describes the relation between physical
quantities measured by two observers moving at constant speed with
respect to each other \cite{Jackson75}.  Electric and magnetic fields
mix under the Lorentz transformation, i.e.~we need to calculate both
$\bE(\bbr,t)$ and $\bB(\bbr,t)$ for obtaining $\bE'(\bbr',t')$.  
More precisely, the strategy will be the following:
\begin{itemize}
\item Calculate $\bE(\bbr,t)$ and
  $\bB(\bbr,t)$ generated by $\bj(\bbr,t)$
  for every space-time point in $\mathcal{R}$. 
\item Apply a Lorentz transformation of these two fields in order to
  get $\bE'(\bbr,t)$, the electric field in the frame
  $\mathcal{R}'$ for every space-time point in $\mathcal{R}$.
\item Calculate the four-vector $(ct',\bbr')$ in the moving
  frame by Lorentz transforming $(ct,\bbr)$ in order to have
  $\bE'(\bbr',t')=\bE'(LT(\bbr),LT(t))$ in
  $\mathcal{R}'$. 
\end{itemize}
We neglect the general relativistic effects due to
the acceleration on the elliptic orbit of the satellite and assume
that the satellite flies with constant speed on a straight line 
over a plane.\\ 

For a given time dependent charge density $\rho(\bbr,t)$ and current density
$\bj(\bbr,t)$, the exact expressions for $\bE(\bbr,t)$ and
$\bB(\bbr,t)$ are (see eqs.(6.55,6.56) in 5th German edition of
\cite{Jackson75}) 
\begin{eqnarray}
\label{elec_field1}
 \bE(\bbr,t)&=&\frac{1}{4\pi\epsilon_0}\int d^3r''
 \Big\{
 \frac{\hat{R}}{R^2}\rho(\bbr'',t-R/c)+
   \frac{\hat{R}}{cR} 
\frac{\partial\rho(\bbr'',t-R/c)}{\partial t} \nonumber \\
 &&-\frac{1}{c^2 R}
 \frac{\bj(\bbr'',t-\frac{R}{c})}{\partial t}
 \Big\},\\ 
\label{magn_field1}
 \bB(\bbr,t)&=&-\frac{\mu_{0}}{4\pi}\int d^3r''
\frac{\hat{R}}{R}\times
\bigg(\frac{1}{R}\bj(\bbr'',t-\frac{R}{c})+\frac{1}{c}
\frac{\partial\bj(\bbr'',t-\frac{R}{c})}{\partial t}
\bigg),
\end{eqnarray}
where $\bR=\bbr-\bbr''$, $R=|\bR|$,  
$\hat{R}/R$ is the unit vector in $(\bbr-\bbr'')$-direction, and
$\epsilon_0,\mu_{0}$ are 
the electric susceptibility and magnetic permeability of vacuum,
respectively. Eqs.(\ref{elec_field1},\ref{magn_field1}) show that the
electric and magnetic fields are fully retarded with respect to the
sources by the propagation time $R/c$.  Both fields contain terms
present also for 
static charges and currents that decay according to the Coulomb-law with
distance, i.e.~as $1/R^2$. However, in the far field $R\gg \lambda$,
where $\lambda$ is the wavelength, these terms are
negligible compared to the ones decaying as $1/R$ and which arise only
for time-dependent charge and current densities
\cite{nanzer_microwave_2012}.  In the following we 
will take into account only these radiation fields. In the far-field
approximation, the fields are 
given by
\begin{eqnarray}
\label{elec_field1b}
 \bE(\bbr,t)&=&\frac{1}{4\pi\epsilon_0}\int d^3r''
 \Big\{
   \frac{\hat{R}}{cR} 
\frac{\partial\rho(\bbr'',t-R/c)}{\partial t}-\frac{1}{c^2 R}
 \frac{\bj(\bbr'',t-\frac{R}{c})}{\partial t}
 \Big\},\\ 
 \label{magn_field1b}
 \bB(\bbr,t)&=&-\frac{\mu_{0}}{4\pi}\int d^3r''
\frac{\hat{R}}{R}\times\frac{1}{c} \partial_{t}\bj(\bbr'',t-\frac{R}{c}), 
\end{eqnarray}

The Lorentz transformation of the electric and magnetic fields from
$\mathcal{R}$ to $\mathcal{R}'$ reads
\cite{Jackson75}, 
\begin{eqnarray}
\label{elec_field2}
\bE'&=&\gamma(\bE+c \bbeta \times \bB) - \frac{\gamma^{2}}{\gamma+1} \bbeta(\bbeta\cdot \bE), \\
\label{magn_field2}
\bB'&=&\gamma(\bB-\bbeta \times
\bE/c)-\frac{\gamma^{2}}{\gamma+1}
\bbeta(\bbeta\cdot \bB), 
\end{eqnarray}
where $\bbeta=\bv_{s}/c$, $\beta =
|\bbeta|$ and $\gamma = 1/\sqrt{1-(v_{s}/c)^{2}}$. We deduce
the expression of the electric field in $\mathcal{R}'$ as function of
space-time coordinates in $\mathcal{R}$,  
\begin{eqnarray}
\label{elec_field3}
\bE'(\bbr,t)&=&-\frac{\gamma\mu_{0}}{4\pi} \int
\frac{d^3r''}{R} 
\bigg[\partial_{t}(\bj-c\hat{R}\rho)+
(\bbeta\cdot\partial_{t}\bj)\hat{R} -(\bbeta \cdot
\hat{R})\partial_{t} \bj \nonumber \\
&&-\frac{\gamma}{\gamma+1}
\bbeta\cdot \partial_{t}(\bj-c\hat{R}\rho)\bbeta
\bigg], 
\end{eqnarray}
where the arguments of $\rho$ and $\bj^{t}$ are
$(\bbr'',t-\frac{R}{c})$.\\

The next step consists of Lorentz transforming the space-time coordinates in $\mathcal{R}$
in order to get the expressions of $(\bbr,t)$ as function of
$(\bbr',t')$,
\begin{eqnarray}
\label{t}
t &=& \gamma (t'+\frac{\bbeta\cdot \bbr'}{c}), \\
\label{r}
\bbr &=& \bbr' + \frac{\gamma -1}{\beta^{2}}(\bbr'\cdot \bbeta)\bbeta+\gamma \bbeta ct'.
\end{eqnarray}
Inserting (\ref{t},\ref{r}) in (\ref{elec_field3}), one finds that
$\bE'(\bbr',t')$ is still given by (\ref{elec_field3}),
where, however, the arguments of $\rho$ and $\bj^{t}$, to be
inserted after differentiation with respect to $t$, are
now 
$\big(\bbr'',\gamma(t'+\frac{\bbeta\cdot\bbr'}{c})-\frac{R}{c}\big)$,
and $\bbr$ is to be replaced everywhere on the right hand side,
including in $\bR$, $R$, and $\hat{R}$, by 
expression (\ref{r}).

\subsection{Expansion in $\beta$}
Eq.~(\ref{elec_field3}) can still be further simplified if we take
into account that $\beta$ is typically very small. The example
of SMOS with $h\simeq 700$\,km,
$v_s\simeq 7$\,km/s, gives $\beta \sim v_{s}/c \sim 10^{-5}$. It makes
therefore sense to expand in powers of $\beta$.  We keep in the
following systematically terms up to order $\beta$ in all phases, but
neglect corrections of order $\beta$ in the prefactor.  This is
justified by the fact that in the end we calculate correlation
functions, where the phase information is crucial, whereas the overall
amplitudes are irrelevant, as they are, in practice, always
renormalized with respect to the total intensity.  \\
To first order in $\beta$, we have $\gamma  
\simeq 1$ and the following approximations 
\begin{eqnarray}
\label{t2}
t &\simeq&  t' +\frac{\bbeta\cdot \bbr'}{c} \\
\label{R2}
\bbr &\simeq&  \bbr'+ \bv_{s}t' \Rightarrow R  \simeq |\bbr'+\bv_{s}t'-\bbr''|
\end{eqnarray}
With this, one finds 
\begin{equation}
\label{elec_field5}
\bE'(\bbr',t') \simeq -\frac{\mu_{0}}{4\pi} \int
\frac{d^3r''}{R}
\partial_{t}\bigg[\bJ(\bbr'',t)\bigg]_{t=t'+\frac{
    \bbeta\cdot \bbr'-R}{c}}, 
\end{equation}
where we have introduced a total current source
$\bJ(\bbr'',t)=\bj(\bbr'',t)-\hat{R}c\rho(\bbr'',t)$ and 
$R=|\bbr'-\bbr''+v_st'|$. If the microscopic charge density vanishes,
$\rho=0$, then the continuity equation ${\rm div}\bj+\partial_t\rho=0$
implies ${\rm div}\bj=0$. In this case, the current density is purely
transverse, $\bJ=\bj=\bj^t$, where $\bj^t$ denotes the part with
vanishing divergence. In general, 
however, fluctuating microscopic charges can temporarily accumulate,
in which case the part $\partial_t\hat{R}c\rho$ may rapidly dominate
over $\partial_t\bj$, as $\bj=\rho\bv$ for charges moving with a speed
$\bv$, and typically $v\ll c$. For the further derivation, there is no
need to distinguish the two cases, and we therefore keep working with
the total current density $\bJ$. \\
Another approximation is in order: We will be interested only in times
$t'$ smaller or equal to the maximum possible averaging time. As
explained below, this time is set by the time of overflight of a
single pixel, assumed to be at constant temperature.  With a pixel
dimension $\Delta x''\sim \Delta y''\sim$ some 1-10km, we have
$v_{s}t/|\bbr'-\bbr''|\le \Delta x''/h \ll  1$. For the
typical values of SMOS, the maximum $t'$ would be of order 1s, but the
inequality shows that even 
substantially longer times  ($t\lesssim 100s$) can still be accommodated in
the range of validity of the following approximation: 
\begin{eqnarray}
\label{R3}
R &\simeq& \sqrt{(\bbr'+\bv_{s}t'-\bbr'')^{2}} \nonumber \\
&=& |\bbr'-\bbr''|\sqrt{1+2\frac{(\bbr'-\bbr'')\cdot \bv_{s}t'}{|\bbr'-\bbr''|^{2}}+\big(\frac{\bv_{s}t'}{|\bbr'-\bbr''|}\big)^{2}} \nonumber \\
&\simeq& |\bbr'-\bbr''| \bigg(1+\frac{(\bbr'-\bbr'')\cdot \bv_{s}t'}{|\bbr'-\bbr''|^{2}} \bigg) \nonumber \\
&=& |\bbr'-\bbr''|+\hat{e}_{\bbr'-\bbr''}\cdot \bv_{s}t'.
\end{eqnarray}
As for the shift $\bbeta\cdot \bbr'/c$ in the time argument, we
have, with antennae at positions $(x_i',y_i',h)$ and
$|x_i'|,|y_i'|\simeq$ (10-100)m $\ll v_st'$, 
$|\bbeta\cdot \bbr'/c| \ll  \beta^2t'$ for almost all $t'$,
such that this term can also be neglected.  Also note that this term
vanishes exactly for antennae located on a line perpendicular to the
displacement of the satellite, $\bbeta\cdot \bbr'=0$. \\

By neglecting again corrections of the amplitude of order $\beta$ or
higher, the expression of the received electric field at position
$\bbr'$ of the antenna relative to the satellite, expressed in the frame
$\mathcal{R}'$, becomes 
\begin{equation}
\label{elec_field6}
\bE'(\bbr',t') \simeq -\frac{\mu_{0}}{4\pi} \int
\frac{d^3r''}{|\bbr'-\bbr''|} \partial_{t}\bJ(\bbr'',t)
\left.\right|_{t=t'-\frac{R(t')}{c}}, 
\end{equation}
with $R(t')=|\bbr'+\bv_{s}t'-\bbr''|$. 
Analyzing expression (\ref{elec_field6}), one realizes that it could
have been obtained in a naive way by simply replacing $\bbr$ in
the far field expression of $\bE(\bbr,t)$ with the position of the
moving satellite as seen in ${\cal  R}$, and keeping the time
$t$. However, even the meaning of 
the partial time-derivative would have then remained ambiguous: does it
apply to $R(t)$ or not? In the formal, rigorous derivation that we
have followed here, it is clear, that it does not: one first
differentiates $\bJ(\bbr'',t)$ with respect to $t$ and then replaces
$t$ as indicated.  Also, the sequence of well-controlled
approximations discussed above opens the way to systematically deriving
higher-order corrections, which is clearly not possible in the
mentioned ``naive'' approach.

\subsection{Fourier transform}
Thermal sources at the surface of Earth, which are the origin of the
thermal noise detected at the position of the satellite, are
assimilated to random fluctuations as function of time. They can be
expressed through their spectrum $\tilde{\bJ}$, 
\begin{equation}
\label{j_t}
\bJ(\bbr'',t) = \frac{1}{\sqrt{2\pi}} \int_{-\infty}^{\infty} d\omega' e^{i\omega' t}\tilde{\bJ}(\bbr'',\omega').
\end{equation}
Inserting (\ref{j_t}) in (\ref{elec_field6}) yields
\begin{equation}
\label{elec_field7}
\bE'(\bbr',t') \simeq -\frac{\mu_{0}}{4\pi \sqrt{2\pi}} \int \frac{d^3r''}{|\bbr'-\bbr''|} \int d\omega' i\omega' e^{i\omega'(t'-\frac{R(t')}{c})}\tilde{\bJ}(\bbr'',\omega').
\end{equation}
The spectrum of the electric field received at the position of the satellite corresponds to the Fourier transform of (\ref{elec_field7}),
\begin{equation}
\label{elec_field8}
\tilde{\bE}'(\bbr',\omega) \simeq -\frac{\mu_{0}}{8\pi^{2}} \int dt' e^{-i\omega t'} \int \frac{d^3r''}{|\bbr'-\bbr''|} \int d\omega' i\omega' e^{i\omega'(t'-\frac{R(t')}{c})}\tilde{\bJ}(\bbr'',\omega'),
\end{equation}
and using the approximation (\ref{R3}), the previous expression becomes
\begin{eqnarray}
\label{elec_field9}
\tilde{\bE}'(\bbr',\omega) &\simeq& -\frac{\mu_{0}}{8\pi^{2}}  \int \frac{d^3r''}{|\bbr'-\bbr''|} \int d\omega' i\omega' e^{-i\omega' \frac{|\bbr'-\bbr''|}{c}} \tilde{\bJ}(\bbr'',\omega') \nonumber \\
&&\times  \int dt' e^{-i\omega t'} e^{i\omega' (1-\hat{e}_{\bbr'-\bbr''} \cdot \bbeta)t'}.
\end{eqnarray}
By noticing that the only remaining time dependence is in the
exponent, the integral over time variable $t'$ can thus be simplified
to 
\begin{eqnarray}
\int dt' e^{-i\omega t'} e^{i\omega' (1-\hat{e}_{\bbr'-\bbr''} \cdot \bbeta)t'}
&=& 2\pi \delta \big(-\omega 
+\omega'(1-\hat{e}_{\bbr'-\bbr''} \cdot \bbeta) \big) \nonumber \\ 
&=& \frac{2\pi}{|1-\hat{e}_{\bbr'-\bbr''} \cdot \bbeta|} \delta \big(\omega'
-\omega(1-\hat{e}_{\bbr'-\bbr''} \cdot
\bbeta)^{-1} \big)\nonumber\\
&\simeq &2\pi|1+\hat{e}_{\bbr'-\bbr''} \cdot
\bbeta| \delta \big(\omega' 
-\omega(1+\hat{e}_{\bbr'-\bbr''} \cdot \bbeta) \big)\,,\nonumber
\end{eqnarray}
where the last expression is once more correct to order $\beta$. 
By neglecting again corrections of the
electric field amplitude, the spectrum of the
electric field measured in $\mathcal{R}'$ becomes 
\begin{equation*}
\tilde{\bE}'(\bbr',\omega) \simeq -\frac{\mu_{0}}{4\pi} \int d^3r'' \frac{i\omega e^{-i\omega \frac{|\bbr'-\bbr''|}{c}(1+\hat{e}_{\bbr'-\bbr''} \cdot \bbeta)}}{|\bbr'-\bbr''|}  \tilde{\bJ}\big(\bbr'',\omega(1+\hat{e}_{\bbr'-\bbr''} \cdot \bbeta)\big).\nonumber
\end{equation*}
On board of the satellite, the received electric field typically
passes through a filter with filter function $w(\omega)$.  E.g.~for
SMOS, the spectrum is
filtered to a narrow window corresponding to the allowed band of width
$b=2\pi\times 17$MHz centered at 
$\omega_{0}=2\pi\times 1.4135$GHz. The retrieved temporal electric field at the output
of the filter is expressed through inverse Fourier transforming
$\tilde{\bE}'(\bbr',\omega)$ multiplied with the filter
function, 
\begin{eqnarray}
\label{elec_field11}
\bE'(\bbr',t') &=& \frac{1}{\sqrt{2\pi}}\int d\omega w(\omega) \tilde{\bE}'(\bbr',\omega) e^{i\omega t'} \nonumber \\
&\simeq& -\frac{\mu_{0}}{4\pi} \frac{1}{\sqrt{2\pi}} \int d\omega w(\omega) \int d^3r'' \frac{i\omega}{|\bbr'-\bbr''|}e^{i\omega t'} \nonumber \\
&& \times \tilde{\bJ}\big(\bbr'',\omega(1+\hat{e}_{\bbr'-\bbr''} \cdot \bbeta)\big) e^{-i\omega \frac{|\bbr'-\bbr''|}{c}(1+\hat{e}_{\bbr'-\bbr''} \cdot \bbeta)}.
\end{eqnarray}
Since the integration over the frequency variable $\omega$ is from
$-\infty$ to $+\infty$, one can apply the change of variables
$\omega(1+\hat{e}_{\bbr'-\bbr''} \cdot \bbeta)
\rightarrow \omega$. Neglecting once more corrections of the amplitude
of order $\beta$ or higher, and corrections of the phase of order
$\beta^2$ or higher, this leads to 
\begin{eqnarray}
\label{elec_field12}
\bE'(\bbr',t') &\simeq& -\frac{\mu_{0}}{4\pi} \frac{1}{\sqrt{2\pi}} \int i\omega d\omega  \int d^3r'' \tilde{\bJ}(\bbr'',\omega) \nonumber \\
&& \times  \frac{ w\big(\omega(1-\hat{e}_{\bbr'-\bbr''} \cdot \bbeta)\big)}{|\bbr'-\bbr''|} e^{i\omega ( t'-\frac{|\bbr'-\bbr''|}{c})}   e^{-i\omega t' \hat{e}_{\bbr'-\bbr''} \cdot \bbeta}.
\end{eqnarray}
As it was to be expected, the satellite motion generates a Doppler
shift of the frequency of the sources. The Doppler shift appears in
the expression of the filtered electric field by means of a shift of
the frequency of the filter function and the appearance of the phase
factor $e^{-i\omega t' \hat{e}_{\bbr'-\bbr''} \cdot
  \bbeta}$. The Doppler shift to linear order in $\beta$ is
clearly a purely longitudinal one, as is well-known. A transverse
(i.e.~purely relativistic)
Doppler shift would appear at second order in $\beta$. 

\subsection{Correlation function}
We define the correlation (also called the visibility function) of
the electric fields in the frame $\mathcal{R}'$ fixed to the satellite
as 
\begin{equation}
\label{corr_func}
C(\bbr_{1}',t_{1}',\bbr_{2}',t_{2}')\equiv \big<\bE'(\bbr_{1}',t_{1}')\bE'^{*}(\bbr_{2}',t_{2}') \big>.
\end{equation}
The thermal sources at the surface of Earth can be modeled by
Gaussian stochastic processes that are uncorrelated for different
frequencies and positions
\cite{rytov_theory_1959,sharkov_passive_2003},
\begin{equation}
\label{FDT_theo}
\big<\tilde{\bJ}\big(\bbr_{1}'',\omega_{1} \big) \tilde{\bJ}^{*}\big(\bbr_{2}'',\omega_{2} \big) \big> =\frac{l_{c}^{3}}{T_{c}} \delta(\bbr_{1}''-\bbr_{2}'') \delta(\omega_{1}-\omega_{2}) \big<|\tilde{\bJ}\big(\bbr_{1}'',\omega_{1} \big)|^{2}\big>,
\end{equation}
where $l_{c}$ and $T_{c}$ refer to the correlation length and the
correlation time, respectively. \\ 
The averages in (\ref{corr_func}) and (\ref{FDT_theo}) are in
principle over an ensemble of different realizations of the noise
processes, but, assuming ergodicity, they may be replaced by a time
average. The averaging time should be as long as possible to reduce
the fluctuations of the average, but sufficiently short for not mixing
different inequivalent ensembles. In our case this means that the
averaging time should be comparable to the time it takes for the
satellite to fly over one pixel (with assumed constant
temperature). This renders the definition (\ref{corr_func}) operational for a
single pass of the satellite.\\ 
\\
Let $\Delta \bbr=\bbr_{2}'-\bbr_{1}'$ and $\Delta t =
t_{2}'-t_{1}'$. To first order in $|\Delta
\bbr|/|\bbr_{1}'-\bbr''| \sim  10^{-4}$, we have the
following approximations  
\begin{eqnarray}
\label{approx1}
|\bbr_{2}'-\bbr''| &\simeq& |\bbr_{1}'-\bbr''|+\Delta \bbr \cdot \hat{e}_{\bbr_{1}'-\bbr''}, \\
\label{approx2}
\hat{e}_{\bbr_{2}'-\bbr''} \cdot \bbeta &\simeq& \hat{e}_{\bbr_{1}'-\bbr''} \cdot \bbeta.
\end{eqnarray}
Finally, using (\ref{elec_field9}), (\ref{FDT_theo}), (\ref{approx1}) and (\ref{approx2}), the expression (\ref{corr_func}) of the correlation function becomes
\begin{eqnarray}
\label{corr_func_1}
C(\bbr_{1}',t_{1}',\bbr_{2}',t_{2}') &\simeq& \frac{l_{c}^{3}}{2\pi T_{c}} (\frac{\mu_{0}}{4\pi})^{2} \int d\omega \omega^{2} \int d^3r'' \big<\big|\tilde{\bJ}\big(\bbr'',\omega\big)\big|^{2}\big>\\
&& \times \frac{ \big|w\big(\omega(1-\hat{e}_{\bbr_{1}'-\bbr''} \cdot \bbeta)\big)\big|^{2}}{|\bbr_{1}'-\bbr''||\bbr_{2}'-\bbr''|} \exp \big[-i\omega \Delta t + i\frac{\omega}{c}(\Delta \bbr+\Delta t \bv_{s})\cdot \hat{e}_{\bbr_{1}'-\bbr''}\big].\nonumber
\end{eqnarray}  
We assume that 
$\omega^{2}\big<\big|\tilde{\bJ}\big(\bbr'',\omega\big)\big|^{2}\big>$
which is related to the brilliance temperature depends only weakly on
frequency (compared to the rapid oscillations of the phase as function
of $\omega$) over the bandwidth $b$, $\omega^{2}\big<\big|\tilde{\bJ}\big(\bbr'',\omega\big)\big|^{2}\big> \simeq \omega_{0}^{2}\big<\big|\tilde{\bJ}\big(\bbr'',\omega_{0} \big)\big|^{2}\big>$.  It is then convenient to invert
the applied change of variables  
$\omega(1-\hat{e}_{\bbr_{1}'-\bbr''} \cdot
\bbeta) \rightarrow \omega$, and one easily finds 
\begin{eqnarray}
\label{corr_func_2}
C(\bbr_{1}',t_{1}',\bbr_{2}',t_{2}') &\simeq& \frac{l_{c}^{3}}{2\pi T_{c}} (\frac{\mu_{0}}{4\pi})^{2} \int d\omega \omega_{0}^{2} |w(\omega)|^{2} \int d^3r''  \frac{\big<\big|\tilde{\bJ}\big(\bbr'',\omega_{0}\big)\big|^{2}\big> }{|\bbr_{1}'-\bbr''||\bbr_{2}'-\bbr''|}\\
&& \times \exp \big[-i\omega \Delta t - i\omega \Delta t\hat{e}_{\bbr_{1}'-\bbr''}\cdot \bbeta + i\frac{\omega}{c}(\Delta \bbr+\Delta t \bv_{s})\cdot \hat{e}_{\bbr_{1}'-\bbr''}\big]. \nonumber 
\end{eqnarray}
By noticing that $\bbeta=\bv_{s}/c$ and neglecting
once more corrections of order $\beta^2$ in the phase, it follows that
\begin{eqnarray}
\label{corr_func_3}
C(\bbr_{1}',t_{1}',\bbr_{2}',t_{2}') &\simeq& \frac{l_{c}^{3}}{2\pi T_{c}} (\frac{\mu_{0}}{4\pi})^{2} \int d\omega \omega_{0}^{2} |w(\omega)|^{2}\\
&&\times   \int d^3r''  \frac{\big<\big|\tilde{\bJ}\big(\bbr'',\omega_{0}\big)\big|^{2}\big> }{|\bbr_{1}'-\bbr''||\bbr_{2}'-\bbr''|} \exp \big[-i\omega \Delta t + i\frac{\omega}{c}\Delta \bbr\cdot \hat{e}_{\bbr_{1}'-\bbr''}\big]. \nonumber 
\end{eqnarray}
We clearly see that the two phases $-\omega\Delta
t\hat{e}_{\bbr'-\bbr''}\cdot \bbeta$ of the Doppler shift and
$(\omega/c)\Delta t\bv_s\cdot\hat{e}_{\bbr'-\bbr''}$ corresponding to
the virtual baseline in the direction of the motion of the satellite
cancel. \\
\\
Finally, by considering a simple rectangular filter function
of bandwidth $b$, 
$$
w(\omega)= \left\{
\begin{array}{l}
  1 \quad \mbox{for} \>\> \omega_{0}-b/2 \leq \omega \leq \omega_{0}+b/2, \\
  0 \quad \mbox{elsewhere},
\end{array}
\right.$$ 
the integral over $\omega$ can be performed. To first order in
$\beta$,  one finds
\begin{eqnarray}
\label{corr_func_4}
C(\bbr_{1}',t_{1}',\bbr_{2}',t_{2}') &\simeq & K\int  d^3r'' \big<\big|\tilde{\bJ}\big(\bbr'',\omega_{0} \big)\big|^{2}\big>  \nonumber \\
&&\times   \frac{\exp \big[i\omega_0\big(- \Delta t + \frac{1}{c}\Delta \bbr\cdot \hat{e}_{\bbr_{1}'-\bbr''}\big) \big]}{|\bbr_{1}'-\bbr''||\bbr_{2}'-\bbr''|}   \nonumber \\
&&\times \mbox{sinc} \big[\frac{b}{2}\big(- \Delta t + \frac{1}{c}\Delta \bbr\cdot \hat{e}_{\bbr_{1}'-\bbr''} \big) \big]. 
\end{eqnarray}
where $\mbox{sinc}(x) \equiv \mbox{sin}(x)/x$, and the constant $K=l_{c}^{3}b
\omega_{0}^{2}\mu_{0}^2/(32\pi^3 T_{c})$. 
Eq.(\ref{corr_func_4}) is
our main result. It generalizes the Van Cittert--Zernike theorem to an
observer moving with respect to the sources and to a finite
time-interval $\Delta t$ between the measurements of the electric
fields, as we discuss now.

\section{Discussion}
A passive micro-wave interferometer for Earth observation measures the complex spatial
correlation field, or the visibility function, of the incident
electric field originating from  thermally fluctuating sources on
Earth's  surface. The  
Van Cittert--Zernike theorem describes the Fourier transform
relationship between a spatial intensity distribution of these incoherent
sources of radiation and its associated
visibility function. In our notation, the theorem can be written as
\begin{equation}
C_{VCZ}(\bbr_{1}',t_{1}',\bbr_{2}',t_{2}') \simeq K\int  d^3r'' \big<\big|\tilde{\bJ}\big(\bbr'',\omega_{0} \big)\big|^{2}\big>  \frac{\exp \big[i\omega_0\big(  \frac{1}{c}\Delta
    \bbr\cdot \hat{e}_{\bbr_{1}'-\bbr''}\big)
  \big]}{|\bbr_{1}'-\bbr''||\bbr_{2}'-\bbr''|}. \label{vCZ}
\end{equation}
It shows that the set of equal time visibility functions obtained
from different 
antennae pairs is given by the spatial 2D Fourier transform of the intensity
distribution of the sources, where the phases of the Fourier transform
are the scalar products of the wave-vector from the source to an
antenna and the vector joining two
antennae. \\

In the standard derivation of the theorem, sources and observer are
taken at rest with respect to each other, and only electric fields
observed at the same time (with respect to the observer
reference--frame) but at different positions are correlated.  We
recover the 
standard form of the VCZT when setting  $\Delta t=0$, and 
considering a small bandwidth, $b\Delta \bbr/c\ll 1$. The latter condition
allows one to approximate the ${\rm sinc}$-function by $\mbox{sinc}(x)\simeq
1$. 
The speed of the satellite $v_s$ has disappeared
from the expression 
already in (\ref{corr_func_3}) with the cancellation of the Doppler
shift and the phase related to the virtual baseline created by the
displacement of the satellite during time $\Delta t$. Thus, {\em the
standard form (\ref{vCZ}) is valid also at finite speed to first order
  in} $\beta$.  
Our study therefore extends the validity of the standard form (\ref{vCZ}) of
the Van Cittert--Zernike theorem to the case of an observer moving
with constant speed with respect to the sources. \\ 

The cancellation of the phases due to the Doppler shift and
the virtual baseline shows that it is not possible to create a
temporal aperture synthesis by correlating the observed time-dependent
electric 
fields delayed by the travel time of the
satellite in the direction of the virtual baseline. 
In addition, a  finite time interval $\Delta t$ between two
observations leads to {\em i.)} a strong suppression of the amplitude
of the 
correlation function and {\em ii.)} a rapidly oscillating phase
factor.  The first effect results from the sinc-term.  If $\Delta t\sim
\Delta\bbr/v_s$, the term due to $\Delta t$ in the argument of the
sinc is up to a factor $c/v_s$ larger than the second one. The second
one, on the other hand, has to be of order one if the spatial aperture
in the direction of the real baseline $\Delta r$ is supposed to
work. Therefore, the amplitude of the correlation function is
suppressed by a factor
$\sim v_s/c$ relative to the standard case  with $\Delta t=0$. The
rapidly oscillating phase factor is given by 
$\exp(-i\omega_0\Delta t)$.  This phase overwhelms the
information in the cross-track direction contained in the phase
$(\omega_0/c)\Delta \bbr\cdot
\hat{e}_{\bbr_{1}'-\bbr''}$.  

Thus, while
the hope of being able to use the virtual baseline created by a moving
satellite for imaging purposes through correlating the time-dependent
fields  shifted only by the travel time over the virtual baseline is
disappointed,  our derivation justifies the neglect of the 
Doppler effect in existing satellite-based passive radiometers based
on the standard Van Cittert--Zernike theorem \cite{SMOS}. 

\section{Conclusion}
We have examined the possibility of temporal aperture synthesis for
satellite-based  
passive microwave observation of Earth, where a virtual baseline is
created by the motion of the satellite in order to enhance the spatial
resolution.  Our study shows that  the interesting phase
information in the along-track 
direction obtained from a time shift of the
fields corresponding to the travel
time over the virtual baseline is exactly canceled by the
first order (longitudinal) Doppler effect. Furthermore, the time shift
yields a 
large uncompensated frequency-dependent phase 
overwhelming the information in the cross-track direction, and a
drastic reduction of the amplitude of the correlation function. 
Therefore, by correlating in this way the
time-dependent signals received by a 1D antennae array with
pixel-independent shifts corresponding only to the travel time of the
satellite, one cannot 
reconstruct the brilliance temperature in both along-
and cross-track directions. 
\\ 

Nevertheless, our result (\ref{corr_func_4}) constitutes a generalization of the
Van Cittert--Zernike theorem (\ref{vCZ}) to the case of an observer moving 
with respect to the sources, and the correlation of electric field
measurements at different times. By deriving 
the electric fields in the moving frame from first principles, we have
shown that the longitudinal Doppler effect cancels exactly in the
correlation function. The standard
Van Cittert--Zernike theorem for equal time
correlations therefore holds even for a moving observer with
substantially different Doppler shifts in different directions of
sight.  

\section{Acknowledgments}
The authors are well aware that without the support of the CESBIO SMOS
team and the CNES project management, this work would not have
succeeded. We are in particular very thankful for the technical
support from Fran\c cois Cabot, Eric Anterrieu, Ali Khazaal, Guy
Lesthi\'evant and Yan Soldo as well as Linda Tomasini the PASO project
manager. We have most appreciated the very valuable educated
suggestions from Jean-Michel Morel and Claire Boyer concerning the
spatio-temporal synthesis modeling. \\


\end{document}